\documentstyle[epsf,psfig]{mn}

\newif\ifAMStwofonts

\ifoldfss
  \ifCUPmtlplainloaded \else
    \NewTextAlphabet{textbfit} {cmbxti10} {}
    \NewTextAlphabet{textbfss} {cmssbx10} {}
    \NewMathAlphabet{mathbfit} {cmbxti10} {} 
    \NewMathAlphabet{mathbfss} {cmssbx10} {} 
  \fi
  \ifAMStwofonts
    \ifCUPmtlplainloaded \else
      \NewSymbolFont{upmath} {eurm10}
      \NewSymbolFont{AMSa} {msam10}
      \NewMathSymbol{\upi}     {0}{upmath}{19}
      \NewMathSymbol{\umu}     {0}{upmath}{16}
      \NewMathSymbol{\upartial}{0}{upmath}{40}
      \NewMathSymbol{\leqslant}{3}{AMSa}{36}
      \NewMathSymbol{\geqslant}{3}{AMSa}{3E}

      \let\leq=\leqslant 
      \let\geq=\geqslant 
    \fi
  \fi
\fi 

\ifnfssone
  \newmathalphabet{\mathit}
  \addtoversion{normal}{\mathit}{cmr}{m}{it}
  \addtoversion{bold}{\mathit}{cmr}{bx}{it}
  \newmathalphabet{\mathbfit} 
  \addtoversion{normal}{\mathbfit}{cmr}{bx}{it}
  \addtoversion{bold}{\mathbfit}{cmr}{bx}{it}
  \newmathalphabet{\mathbfss} 
  \addtoversion{normal}{\mathbfss}{cmss}{bx}{n}
  \addtoversion{bold}{\mathbfss}{cmss}{bx}{n}
  \ifAMStwofonts
    \ifCUPmtlplainloaded \else
      \UseAMStwoboldmath
      \makeatletter
      \new@mathgroup\upmath@group
      \define@mathgroup\mv@normal\upmath@group{eur}{m}{n}
      \define@mathgroup\mv@bold\upmath@group{eur}{b}{n}
      \edef\UPM{\hexnumber\upmath@group}
      \new@mathgroup\amsa@group
      \define@mathgroup\mv@normal\amsa@group{msa}{m}{n}
      \define@mathgroup\mv@bold\amsa@group{msa}{m}{n}
      \edef\AMSa{\hexnumber\amsa@group}
      \makeatother
      \mathchardef\upi="0\UPM19
      \mathchardef\umu="0\UPM16
      \mathchardef\upartial="0\UPM40
      \mathchardef\leqslant="3\AMSa36
      \mathchardef\geqslant="3\AMSa3E

      \let\leq=\leqslant 
      \let\geq=\geqslant 
    \fi
  \fi
\fi 

\ifnfsstwo
  \DeclareMathAlphabet{\mathbfit}{OT1}{cmr}{bx}{it}
  \SetMathAlphabet\mathbfit{bold}{OT1}{cmr}{bx}{it}
  \DeclareMathAlphabet{\mathbfss}{OT1}{cmss}{bx}{n}
  \SetMathAlphabet\mathbfss{bold}{OT1}{cmss}{bx}{n}
  \ifAMStwofonts
    \ifCUPmtlplainloaded \else
      \DeclareSymbolFont{UPM}{U}{eur}{m}{n}
      \SetSymbolFont{UPM}{bold}{U}{eur}{b}{n}
      \DeclareSymbolFont{AMSa}{U}{msa}{m}{n}
      \DeclareMathSymbol{\upi}{0}{UPM}{"19}
      \DeclareMathSymbol{\umu}{0}{UPM}{"16}
      \DeclareMathSymbol{\upartial}{0}{UPM}{"40}
      \DeclareMathSymbol{\leqslant}{3}{AMSa}{"36}
      \DeclareMathSymbol{\geqslant}{3}{AMSa}{"3E}

      \let\leq=\leqslant 
      \let\geq=\geqslant 
    \fi
  \fi
\fi 

\ifCUPmtlplainloaded \else
  \ifAMStwofonts \else 
    \def\upi{\pi}
    \def\umu{\mu}
    \def\upartial{\partial}
  \fi
\fi

\newcommand{\kms}{\thinspace\hbox{$\hbox{km}\thinspace\hbox{s}^{-1}$}}
\newcommand{\vsini}{\thinspace\hbox{$v\thinspace\hbox{sin}i$}}

\newcommand{\ha}{H$\alpha$}

\newcommand{\hei}{He\,{\sc i}}

\newcommand{\deltav}{$\Delta V_{peak}$}

\newcommand{\ewha}{$W_{\lambda H\alpha}$}
\newcommand{\ewhei}{$W_{\lambda He I}$}
\newcommand{\narrowto}{\hspace{-0.02in}\to\hspace{-0.01in}}

\title{Cyclical behaviour and disc truncation in the Be/X-ray binary A0535+26}

\author[N.J.Haigh, M.J.Coe, J.Fabregat]
       {N.J.Haigh$^{1}$, M.J.Coe$^{1}$, J.Fabregat$^{2}$ \\
$^{1}$Department of Physics and Astronomy, The University, Southampton, SO17 1BJ, UK. \\
$^{2}$Observatorio Astron{\' o}mico, Universidad de Valencia, 46100 Burjassot, Spain \\ \\}
\date{Accepted ???. Received ???}

\pagerange{\pageref{firstpage}--\pageref{lastpage}}
\pubyear{2003}

\begin{document}

\maketitle

\label{firstpage}

\begin{abstract}

A0535+26 is shown to display quantised IR excess flux states, which are interpreted as the first observational verification of the resonant truncation scheme proposed by Okazaki and Negueruela (2001) for BeXRBs. The simultaneity of X-ray activity with transitions between these states strongly suggests a broad mechanism for outbursts, in which material lost from the disc during the reduction of truncation radius is accreted by the NS. Furthermore changes between states are shown to be governed by a $\sim$1500 day period, probably due to precession of the Be disc, which profoundly dictates the global behaviour of the system. Such a framework appears to be applicable to BeXRBs in general.

\end{abstract}


\section{Introduction}
\label{section:intro}

Roughly two thirds of High Mass X-ray Binaries possess a Be star as the primary, and are known as Be X-ray binaries or BeXRBs. A neutron star (NS) orbits the O9-B2 primary in $10\to300$days with pronounced eccentricity (e=0.1$\to$0.9). Whilst some systems display only weak persistent X-ray emission (X Per type), most additionally exhibit large transient X-ray outbursts, sometimes close to their Eddington luminosity. Outbursts fall into two types \cite{stella1986}:

(I)  Type I outbursts ($10^{36}-10^{37}erg~s^{-1}$) last a matter of days and occur close to or shortly after periastron.

(II) Type II or 'giant' outbursts of $\sim10^{38}erg~s^{-1}$ occur between orbital phase 0 and 0.5, lasting several tens of days.

Nearly 3 decades after its discovery during a type II event, A0535+26 remains of considerable interest to those studying BeXRBs; due to its proximity high quality data can be obtained with ease, whilst in many ways it appears to be typical of its class. The O9.7IIIe \cite{steele98} primary HDE 245770 displays typical variability in all aspects of spectral, photometric and X-ray properties \cite{clark98,clarklyuty1999,larionov2001}. In addition to type I outbursts which recur at 110-111 day intervals (presumed to be the orbital period $P_{orb}$), `giant' or type II outbursts have occurred less predictably in 1975, 1980, 1989 and 1994. This paper is an attempt to draw together 15 years of multiwaveband variations to compose a coherent global picture of the behaviour of A0535+26.

Integral to the analysis presented here is recent largely theoretical work regarding disc truncation in BeXRBs \cite{natural2001,okazaki2002}. Central to the new picture is the abrupt truncation of the Be disc at a radius smaller than the periastron separation, where the orbital period is resonant with that of the neutron star and tidal interactions exceed viscous forces, the specific resonance being dependant upon the system parameters and the viscosity $\alpha$. The traditional picture of the NS passing through the Be disc is thus erroneous, explaining quiescent X-ray states in systems with robust discs. 

The mechanisms leading to accretion and X-ray outbursts remain unknown except in systems of such high eccentricity that inefficient truncation permits accretion at periastron. In the case of 4U~0115+63 spectroscopy, photometry and X-ray observations have been combined to yield a picture in which quasi-cycles of 3-5 years govern the system \cite{iggy2001,neg2001}. In their interpretation, such cycles commence with disc-loss followed by reformation and expansion until the disc becomes truncated. The disc becomes unstable to warping/tilting, and precession causes alternating single peaked and shell line profiles. Type II outbursts then occur, presumably related to the unstable disc configuration.

This study was limited by the low S/N and sparsity of photometric and spectroscopic data for 4U~0115+63, attributable largely to the profound reddening. The analysis of A0535+26 presented here has the advantage of a 15 year dataset of spectra and IR photometry including many quasi-simultaneous spectral and photometric measurements and the opportunity to build upon the work that has been done in understanding 4U~0115+63.

In Section \ref{section:IRquant} we present the first firm evidence for resonant truncation of the Be disc with the discovery that the IR excess is quantised with relative fluxes matching theoretical predictions. In Section \ref{section:cycles} we discuss a 1500 day cyclical 'looping' behaviour seen in plots of \ewha vs. $m_K$, as truncation radius and emission line flux vary coherently but 90$^\circ$ out of phase.

Section \ref{section:temporal} describes how X-ray outbursts, in particular 'giant' type II outbursts, occur at times of decreasing truncation radius and at a specific cyclical phase. An association of 'expelled' material with NS accretion is strongly suggested and a mechanism for X-ray activity suggested. A further point of interest uncovered in the dataset is the possible IR detection of the NS accretion disc previously inferred from X-ray QPOs \cite{finger1996} during the 1994 type II X-ray event. In Section \ref{section:inclination} data obtained following the 1998 disc-loss event are used in combination with the known disc size to determine the inclination, rotational parameter $\omega$, and the circumstellar disc's radial velocity.

Section \ref{section:periods} discusses the various periodicities observed in A0535+26, their probable origins and mutual associations, in particular highly coherent photometric variations at the oft-detected 103 day period. Thus the phenomenology that has been described regarding 4U~0115+63 is applied to A0535+26 and some modifications suggested. The integral role of the warped, precessing disc invoked in 4U~0115+63 is supported in the case of A0535+26, though in a less transient r{\^ o}le. We are thus led to a closer understanding of the global behaviour of the system.

\section{Observations}
\label{section:obs}

\subsection{Photometry}

\begin{figure}
\begin{center}
\psfig{file=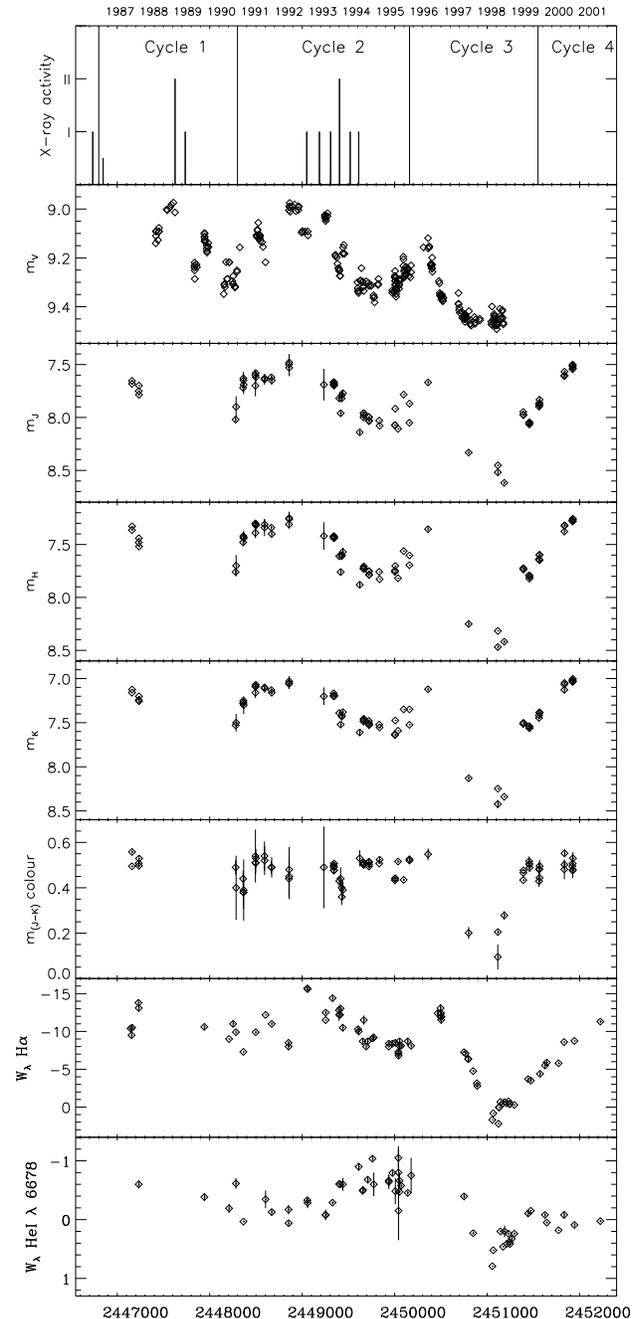,width=3.3in,angle=0}
\caption{Observational parameters of A0535+26 from 1987-2001. The upper panel shows detected X-ray activity and the cycles described in Section \ref{section:cycles}.}
\label{fig:a0535jhk}
\end{center}
\end{figure}

The Southampton/Valencia collaboration has gathered infrared observations at JHK from 1987 to 2001 with the 1.5-m Carlos S\'anchez telescope (TCS), located at the Teide Observatory in Tenerife, Spain. The data were reduced following the procedure described by Manfroid (1993). Instrumental values were transformed to the TCS standard system (Alonso, Arribas \& Mart{\'\i}nez 1998). The data acquired for A0535+26 are plotted in Figure \ref{fig:a0535jhk}.

For the histograms in Figure \ref{fig:a0535phothisto}, multiple observations spanning several days from individual observing runs are represented by single data points, which are thus separated by at least 10 days to prevent statistical bias arising from unrepresentative data sampling. Typical photometric errors are $\pm0.05$.

Photometric $m_{JHK}$ measurements at epoch 17/12/1997 have been extracted from the 2MASS catalogue at IRSA.

The $m_V$ photometric data presented by Lyuty and Za{\u \i}tseva (2000), and available online as a VizieR catalogue, are used here and also plotted in Figure \ref{fig:a0535jhk}. Errors are $\pm0.005$.

\subsection{Spectroscopy}

\begin{table*}
\caption{Unpublished spectroscopic observations of HDE 245770.}
\begin{tabular}{|l|l|l|l|}
\hline
Date      & Telescope           & Instrument & Details\\
\hline
13/12/1991& INT                 & IDS          & EEV5, 500 camera, R1200R\\
8-10/3/1993&SAAO 1.9m           & Spectrograph & RPCS\\
12/2/1995 & JKT                 & AGB          & EEV5, AGBX \\
4/4/1996  & SAAO 1.9m           & Spectrograph & RPCS\\
28/10/1997& JKT                 & AGB          & TEK4, AGBX \\
6/2/1998  & SAAO 1.9m           & Spectrograph & SITe, 1200 l$mm^{-1}$\\
24/12/1998& WHT                 & ISIS blue arm& EEV12, R600B \\
27/3/1999 & WHT                 & ISIS red arm & TEK2, R600R \\
21/9/1999 & INT                 & IDS          & TEK5, 235 camera, R1200R \\
20/10/1999& WHT                 & UES          & SITe1 \\
22/3/2000 & INT                 & IDS          & TEK5, 235 camera, R1200R \\
10/4/2000 & INT                 & IDS          & TEK5, 500 camera, R1200Y \\
15/8/2000 & INT                 & IDS          & TEK5, 500 camera, R1200Y \\
16/10/2000& Mt. Skinakas 1.3m   & Spectrograph & 1302r l$mm^{-1}$ +$80\mu m$ + ISA 608\\
5/2/2001  & INT                 & IDS          & TEK5, 500 camera, R1200Y \\
11/11/2001& SAAO 1.9m           & Spectrograph & SITe, 1200 l$mm^{-1}$\\
\hline
\end{tabular}
\label{tab:a0535specobs}
\end{table*}

Spectra of A0535+26 including \ha~and \hei~$\lambda6678$ have been obtained since 1987 at numerous observatories. Observations not described in Clark et al. \shortcite{clark98} or Haigh et al. \shortcite{haigh99} are listed in Table \ref{tab:a0535specobs}. The JKT, INT and WHT data have been obtained through the ING service programme or the ING Archive.

Reduction took place using {\footnotesize STARLINK} packages, with spectral analysis performed in {\footnotesize DIPSO}.

Twelve additional \ha~equivalent width (\ewha) observations between Jan 1997 and Mar 1998 are taken from Piccioni \shortcite{piccioni1998}. Two further values from 11-12/11/1998 are taken from Giovannelli et al. \shortcite{iauc7293}, with one each from 6 Nov 1998 \cite{lyuty2000} and 29 Jan 2000 (Negueruela, private comm.).

\section{IR excess quantisation and truncation}
\label{section:IRquant}

\begin{figure}
\begin{center}
\psfig{file=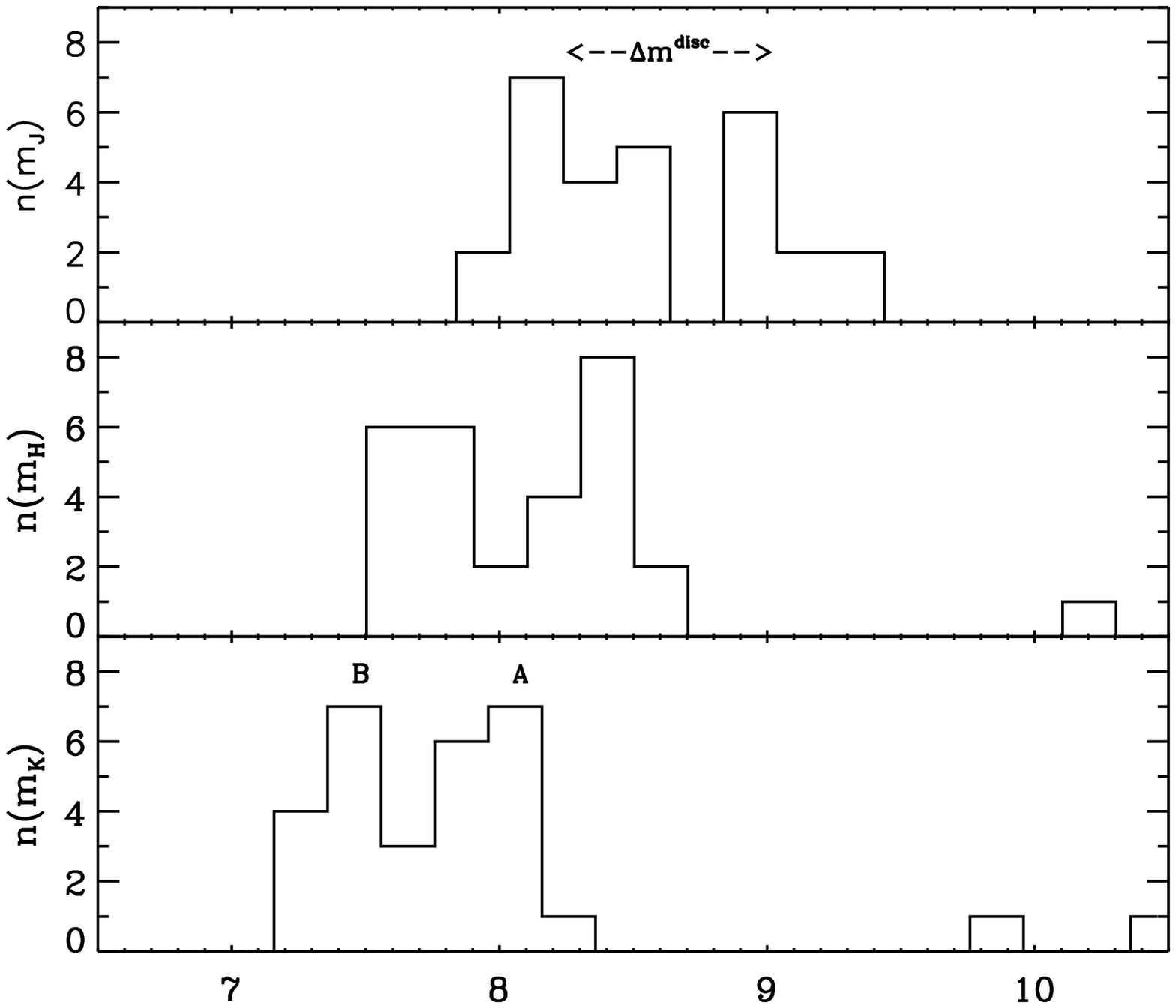,width=3.3in,angle=0}
\psfig{file=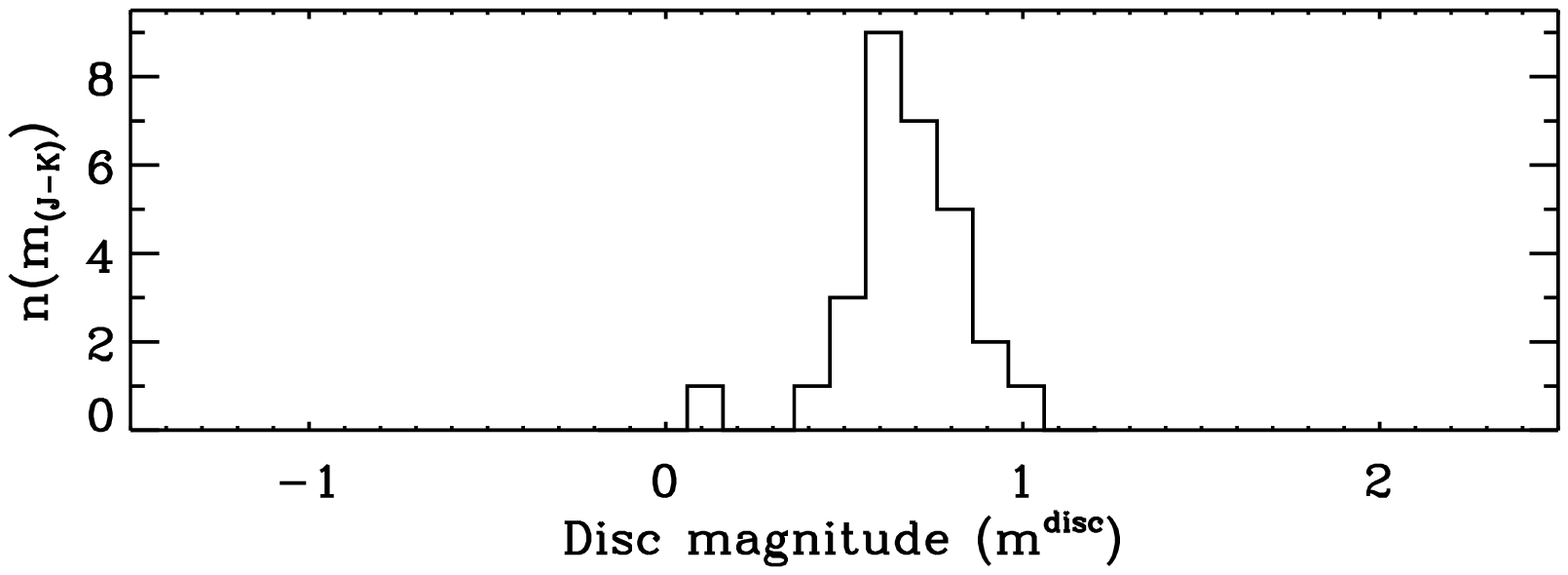,width=3.3in,angle=0}
\caption{Histograms of A0535+26 decretion disc magnitudes $m^{disc}$ in J,H, K and colour index (J-K). Bin widths are 0.2 for $m^{disc}_{JHK}$ and 0.1 for $m^{disc}_{(J-K)}$. Flux states A and B are labelled on the $m^{disc}_K$ histogram.}
\label{fig:a0535phothisto}
\end{center}
\end{figure}

In order to study only the circumstellar decretion disc flux the constant stellar component $m^*$ is subtracted - using stellar magnitudes $m_J^*=8.6$, $m_H^*=8.5$ and $m_K^*=8.5$ - and the resultant magnitude subsequently indicated with the superscript {\em disc}. Note $m_{JHK}^{disc}$ are relatively insensitive to errors in $m_{JHK}^*$.

Figure \ref{fig:a0535phothisto} shows histograms of the $m^{disc}_{JHK,(J-K)}$ photometry obtained from 1987 to 2001. These infrared fluxes display a clear bimodality as the system alternates between faint and bright flux states denoted A and B respectively and separated by $\Delta m^{disc}$ magnitudes. The contrasting constancy of $m_{(J-K)}^{disc}$ shows the dominance of optically thick isothermal emission \cite{dougherty1994}, suggesting that changes in disc area are responsible, rather than optical depth or temperature.

Such a quantized luminosity distribution is a natural consequence of the resonant truncation hypothesis proposed by Okazaki and Negueruela \shortcite{natural2001}. Here tidal torques, which operate most strongly in orbits resonant with that of the NS, truncate the disc at radii meeting the condition $P_{NS}:P_{trunc}=n:1$ where n is an integer. They find that for viscosities of $\alpha=0.013-0.38$ the disc is truncated at the 6:1 resonance, $\alpha=0.038-0.11$ at 5:1 and for $\alpha=0.11-0.40$ the disc is supported out to 4:1.

\begin{figure}
\centerline{\psfig{file=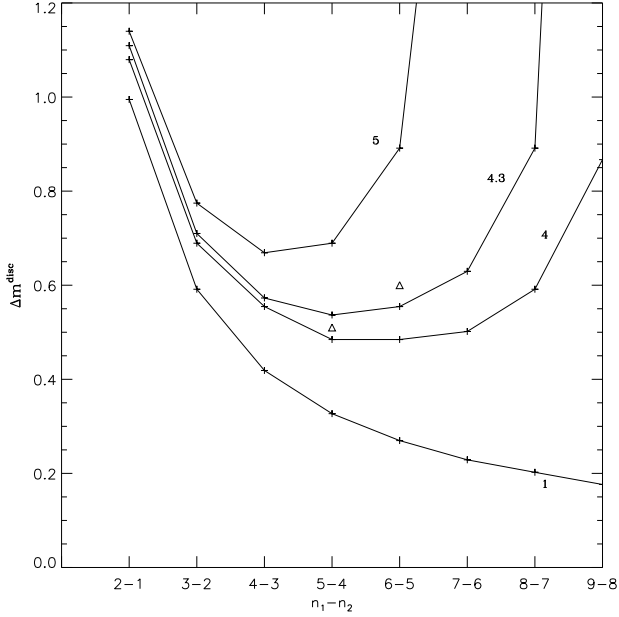,angle=0,width=3.3in}}
\caption{Modelling of disc flux ratios $\Delta m^{disc}$ for changes between disc states truncated at n$_1$:1 and n$_2$:1 resonances. Emission is from an optically thick isothermal disc with central cavity of $r_{inner}=1,4,4.3,5R_*$, where $R_*=1.02\times10^{10}m$ (Vacca et al. 1996). Assuming the identities of states with resonances discussed in the text, the $\triangle$ symbols show the observed values of $\Delta m^{disc}_K$.}
\label{fig:cavity}
\end{figure}

Assuming isothermal optically thick emission, the quantity $\Delta m^{disc}$ is a direct measure of the relative areas of the disc in these two states. Knowing the orbital period of the disc's outer edge at a given resonance, and assuming Keplerian rotation and a non edge-on inclination enables the relative disc radii and thus $\Delta m^{disc}$ to be predicted independently of stellar mass $M_*$. Thus a change in truncation resonance produces a $\Delta m^{disc}$ plotted in Figure \ref{fig:cavity}, dependant on the size of any central cavity, or separation of the inner disc radius from the photosphere.

Rivinius et al. \shortcite{rivinius2001} found that in the typical isolated Be star $\eta~Cen$, mass injection events or 'outbursts' lowered the disc inner radius to $1.2R_*$, but at other times a consistent $3.1\to3.3R_*$ was observed. There is no reason to believe that the same mechanisms should not lead to similar cavities in A0535+26. Curves for different values of $r_{inner}$ assume $R_*=14.7R_\odot$ and $M_*=26M_{\odot}$ \cite{vacca1996} based upon an O9.7IIIe classification; note the star's volume produces a 'cavity' of $r_{inner}=1R_*$. Note that resonances are with the orbital beat frequency of $P_{beat}=103$ days rather than the orbital period of $P_{orb}\simeq111$ days for reasons discussed in Section \ref{section:periods}, though this has a minimal effect.
 
Of the three IR photometric wavebands K has the largest optical depth and most closely approximates the optically thick assumption. Photospheric emission is also minimized in this band, so the $\Delta m_K^{disc}$ data are used hereon, though all wavebands yield similar results. Table \ref{tab:fluxquant} and Figure \ref{fig:a0535phothisto} yield a value for $\Delta m_K^{disc}$ of 0.60; consulting Figure \ref{fig:cavity} this suggests a central cavity $r_{inner}/R_*\simeq 4.3$ taking states A and B to be truncated at 6:1 and 5:1 respectively. Whilst the various assumptions make this only an approximate result, $M_*$ and $R_*$ are sufficiently well determined to yield maximum errors in $r_{inner}/R_*$ of $\sim30\%$ requiring a central cavity to be present. Lack of isothermality will cause $r_{inner}$ to be underestimated if as expected $T_{disc}$ decreases radially. Departures from optical thickness will primarily affect the outer disc and will have the same effect. This result is in encouraging agreement with the central cavities observed in $\eta~Cen$, whereas using radii for dwarves yields $r_{inner}/R_*\simeq 7.5$. This result therefore supports the giant (III) classification.

\begin{table}
\centering
\caption{Observational parameters and {\em ranges} for the two observed flux states from 1987 to 2001, the additional state C proposed from the literature and disc-loss. 1:Using $m_B$ of Lyuty et al. (2000) and $m_{(B-V)}=0.55$, 2: 11-13/12/1976, Persi et al.(1979), 3: Haigh et al. (1999), 4: Vo{\u \i}khanskaya (1980), Baratta et al.(1978).}
\begin{tabular}{|l|l|l|l|l|l|}
\hline
State         &C                 &B            &A            &Discloss\\
\hline
\ewha         &$-17\pm3^4$       &$-12\pm4$    &$-10\pm3$    &$\sim2.5$\\
\ewhei        &        ?         &$-0.2\pm0.3$&$-0.8\pm0.4$&$0.3\pm0.2$\\
$m_V$         &$\sim8.85\pm0.1^1$&$9.10\pm0.1$ &$9.30\pm0.1$ &9.45$^3$\\
$m_J$         &$\sim7.34^2$      &$7.60\pm0.1$ &$8.00\pm0.15$&8.49$^3$\\
$m_H$         &$\sim7.02^2$      &$7.35\pm0.1$ &$7.75\pm0.15$&8.39$^3$\\
$m_K$         &$\sim6.71^2$      &$7.10\pm0.1$ &$7.50\pm0.15$&8.34$^3$\\
$m_K^{disc}$  &$\sim6.94$        &$7.45\pm0.1$ &$8.05\pm0.15$&$\geq10$\\
\hline
\end{tabular} 
\label{tab:fluxquant}
\end{table} 

IR photometry \cite{persi1979}, and the $m_V$ data presented by Lyuty and Za{\u \i}tseva \shortcite{lyuty2000}, suggest that the disc was in an even brighter state (C) from the early 1970s until 1980 (see Table \ref{tab:fluxquant}). Optically thick emission in the IR bands requires this to have been a physically larger disc. As Okazaki and Negueruela \shortcite{natural2001} note, truncation at 4:1 places the outer edge of the disc marginally outside the Be star Roche lobe at periastron, and thus represents the largest stable disc state. The identity of state C with this 4:1 truncation, and thus states B and A with 5:1 and 6:1, is appealing as it explains the prodigious X-ray activity of the system during the 1970s, periastric Roche lobe overflow alone enabling accretion and X-ray activity \cite{natural2001}. Identifying state C with a 3:1 truncation places much of the disc outside the Be star Roche lobe at periastron preventing stability, and requires an unrealistic viscosity \cite{okazaki2002}.

Figure \ref{fig:cavity} shows that a change in truncation between 5:1 and 4:1 (state B$\to$C) will produce a $\Delta m_K^{disc}\approx0.56$ and thus predicts $m_K^{disc}\simeq6.89$ for state C, in excellent agreement with the isolated measurement of $m_K^{disc}=6.94\pm0.03$ in the literature (Table \ref{tab:fluxquant}).

It should be noted here that the strongly shepherded structure of a truncated disc leads to highly stable continuum fluxes - this is discussed in Section \ref{section:periods}.

\section{Cyclical behaviour}
\label{section:cycles}

\begin{figure}
\psfig{file=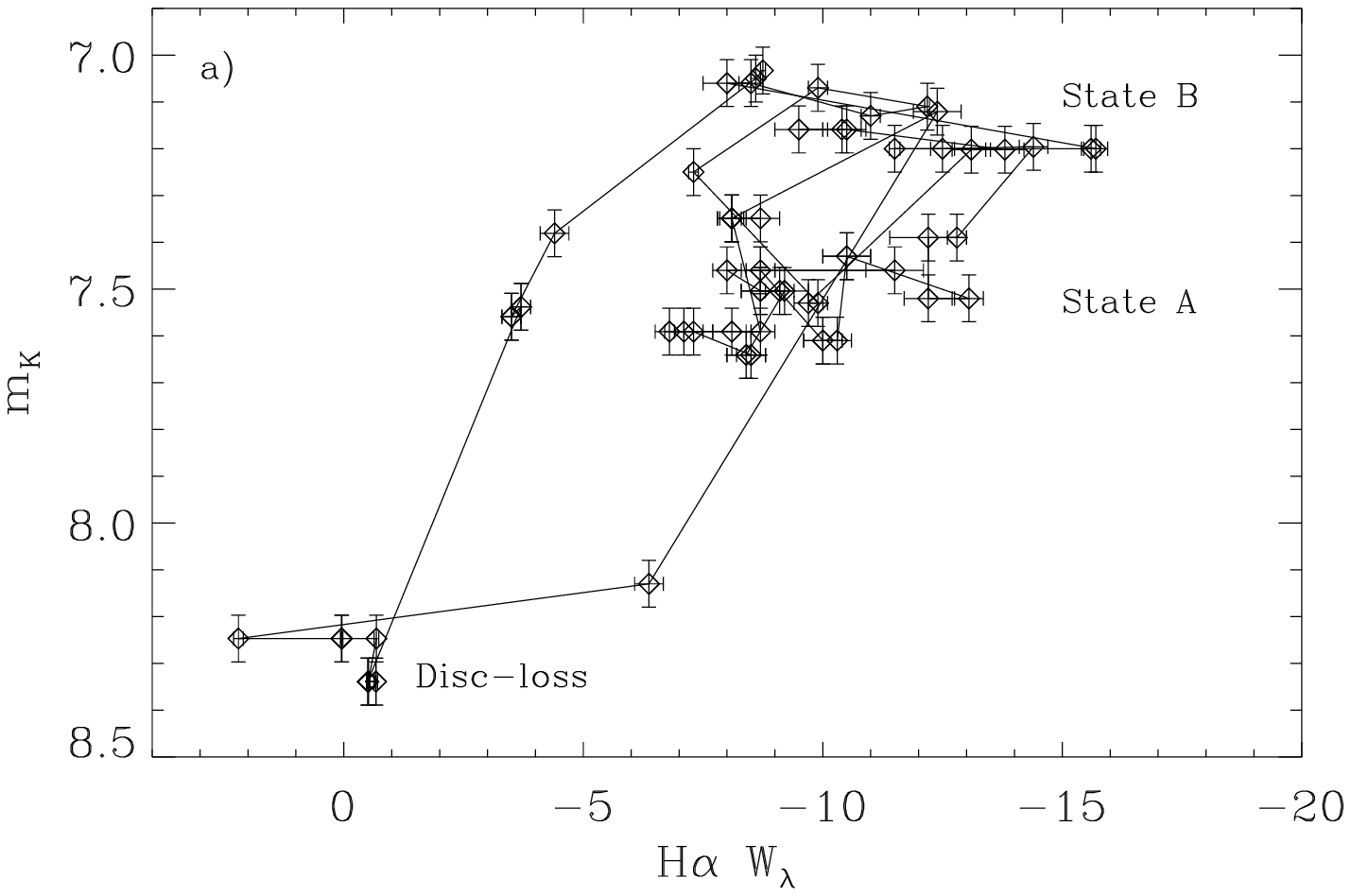,angle=90,width=3.3in}
\psfig{file=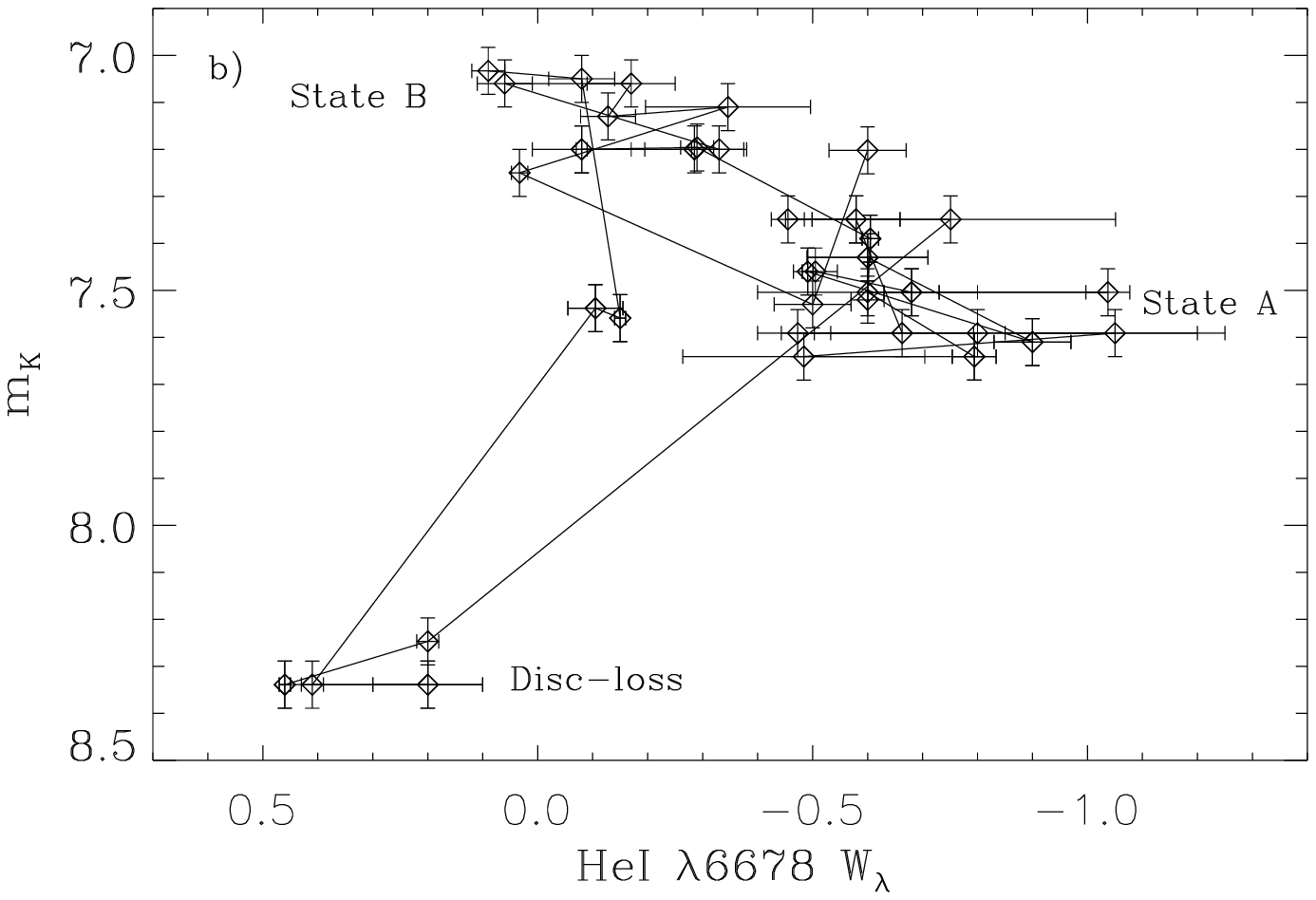,angle=90,width=3.3in}
\caption{Path of $m_K$ vs. \ewha~ (a) and \ewhei$\lambda6678$ (b). Flux states A and B are labelled. In a), movement around the loop is clockwise.}
\label{fig:a0535correlation1}
\end{figure}

Figure \ref{fig:a0535correlation1}a plots $m_K$ vs. \ewha~between 1987 and 2001. For each measurement of \ewha~ the closest photometric $m_K$ measurement - always within 30 days and mostly within 10 - is taken. Flux states A and B are clearly visible as parallel bands of points.

A0535+26 can be seen to follow a clockwise loop with a mean period of $4.36\pm0.5$ years ($1590\pm180$ days) except for those points around the disc-loss episode \cite{haigh99} at lower left. Following Negueruela and Okazaki \shortcite{iggy2001} who observed similar cyclical behaviour in the BeXRB 4U~0115+63, these looping cycles in A0535+26 are defined as commencing following disc-loss, or at the state A$\narrowto$B transition which occurs at the same phase. In terms of the flux states described above, each cycle consists of approximately equal periods first in state B followed by state A. During state B, \ewha~gradually increases before decreasing following the $B\narrowto A$ transition. Figure \ref{fig:oddcolours} shows photometry and X-ray activity during cycle 2, which has the best observational coverage and appears to be typical.

In contrast to \ewha~which varies through a similar range in both states, Figure \ref{fig:a0535correlation1}b shows \ewhei~oscillating between two loci in phase with the \ewha variations. With the state B locus at \ewhei~$\simeq-0.15$ and \ewhei~$\sim-0.60$ in state B, the emission line flux ratio of states A:B$\sim4$, demonstrating that the smaller changes in continuum strength ($\sim1.3$, determined from $m_R$) cannot be responsible. Allowing for infilling of the underlying photospheric absorption \hei$~\lambda6678$ has a $\sim25\%$ greater flux in the fainter state A. This is interpreted as arising from enhanced densities at the small radii where this line is primarily produced \cite{stee1998}, due to the more strongly confined disc in this state. At the same time, the wings of \hei$~\lambda6678$ arising in the inner disc display emission to systematically higher velocities. Thus these diagrams enable the flux state to be determined either from spectra or photometry.

Due to the patchy nature of the dataset, a holistic analysis of optical and IR photometric and spectral data was required to reconstruct the timing and phenomena associated with the ongoing and previous 3 cycles, which are marked at the top of Figure \ref{fig:a0535jhk}. Whilst the $\sim1400$ day photometric component of these cycles has been previously noted in period searches \cite{clarklyuty1999,larionov2001}, the coherent variation of other observational parameters has not. These observations demonstrate that the long period variations observed in A0535+26 are not arbitrary but rather linked to a 1400-1500 day cyclical variation, perhaps precessional, in the system, the nature of which is discussed in Section \ref{section:periods}.

\section{Temporal association with X-ray outbursts}
\label{section:temporal}

When the disc outer radius makes a quantum leap inwards (state C$\to$B or B$\to$A) the material present between those radii must be relocated elsewhere in the system. It is suggestive to note that such a state change was observed in $m_K$ between JD 2449344 (state B) and 2449401 (state A), while a type II X-ray outburst was detected by BATSE \cite{finger1996} from $2449380$ to $2449432$. This simultaneity is well seen in Figure \ref{fig:oddcolours}, and strongly suggests that the giant outburst was the accretion of material previously resident between the 5:1 and 6:1 resonances. The greater duration of variability in $m_V$ relative to $m_K$ is interpreted as being due to the greater optical depth in $m_K$ desensitising the photometry to small variations in disc structure as truncation progressed. The entire 1993-4 series of X-ray outbursts, types I and II, all occurred during the somewhat lengthier truncation process as seen in $m_V$, and presumably correspond to material accreted following removal from the outer disc. Importantly, truncation would appear to be the cause of the X-radiation and not vice-versa, as truncation commences before X-ray emission.

In 1989, precisely one cycle earlier during the state $B\to A$ transition of cycle 1, a type II followed by a type I X-ray outburst occurred in the system. Whilst observations are sparser, this series of events appears extremely similar to the 1993-4 events described above, and we therefore suggest that the same processes were responsible.

Unlike cycles 1 and 2, the type II outburst in Oct. 1980 occurred during state C with $m_V=8.9\pm0.04$ from Jan 1980 until Dec 1980. No state change occurred until some months after the event. However this can be understood, as the 4:1 resonance hypothesised to bound state C is marginally outside the Be star Roche lobe (see Section \ref{section:IRquant}) enabling mass transfer without a change in truncation radius \cite{natural2001}. This may indeed be the mechanism behind many of the X-ray outbursts during the bright photometric state of the 1970s, including the 1975 discovery type II event. As has been noted \cite{lyuty1989} even the quiescent X-ray flux of A0535+26 was elevated from 1975-1980; the identification of this epoch with a 4:1 truncation (state C) permits mass transfer through the inner Lagrangian point at periastron even in the absence of truncation changes. The absence of accurate IR photometry from this period prevents further analysis at the present.

Clark et al. (1999) in discussing the long term photometric variations of A0535+26, noted that X-ray outbursts seemed to occur after a period in which the optical light curve was fading. Clark suggested that this represented an episode in which disc material was lost radially, with X-ray emisison triggered by the interaction of this material with the NS. However, this suggestion was rejected in favour of the traditional 'shell ejection from the primary' scheme.

The question remains of why some transitions lead to X-ray activity while others do not. A possible factor is that though disc truncation may occur at a specific phase in the 1500 day precession cycle, the precession angle changes by $26^\circ$ between each periastron passage so system geometries do not recur precisely. To take a simplistic viewpoint, it may be that whether the NS lies precisely in the disc plane at the time of critical precession phase determines whether a type II outburst occurs or not. The 1996 truncation occurred very early in the cycle, and no X-ray activity was seen.

As with the model of Negueruela et al. \shortcite{neg2001} for 4U~0115+62, this model of X-ray outbursts is effectively independant of Be star activity, and eliminates the unwieldy 'ejections' of material from the Be star frequently invoked as directly fuelling X-ray outbursts \cite{deloore1984}. Such connections between long-term (months-years) photometric changes and X-ray events has not received a great deal of attention in the literature, with much work expecting to observe short optical flares associated with enhanced stellar mass-loss prior to an outburst.

It should be noted here that whilst photometry is unavailable, spectroscopy from 2 Jan 2003 showing \ewha=-13.5\AA~and \ewhei$_{\lambda6678}$=-0.3\AA~ (Laycock, private comm.) confirms that the system is in the latter stages of state B (see Table \ref{tab:fluxquant}). The imminent transition to state A, expected mid 2003, could produce the first X-ray outburst in the A0535+26 system since 1994.

\subsection{IR excess colour changes}
\label{section:disccolourchange}

\begin{figure}
\begin{center}
\centerline{\psfig{file=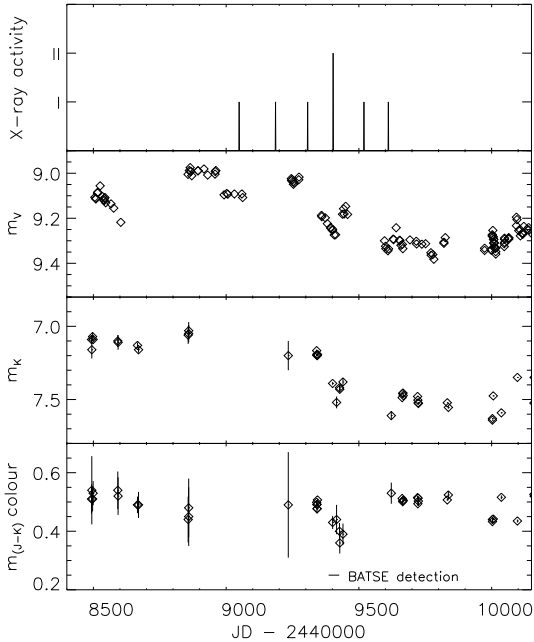,width=3in,angle=0}}
\caption{Photometry throughout cycle 2 including the 1993-4 X-ray outbursts. The duration of the BATSE 20-100 keV detection during the type II event is shown as a short line.}
\label{fig:oddcolours}
\end{center}
\end{figure}

The presence of an accretion disc around the neutron star in A0535+26 has been debated in the literature \cite{giovannelli1990,lyuty2000}. While no evidence has been found in the optical/IR to date, X-ray observations have provided unequivocal evidence during the 1994 giant outburst. Finger et al. \shortcite{finger1996} detected 20-100mHz QPOs and large \.{P} from JD $2449380$ to $2449432$ which they concluded are due to an accretion disc around the NS. Motch \shortcite{motch1991} and Li \shortcite{li97} found that an accretion disc is only present during such giant outbursts, therefore the data obtained at this time presents an excellent opportunity to look for emission from any such disc. 

Figure \ref{fig:oddcolours} plots $m_V$, $m_K$ and the IR colour index $m_{(J-K)}$ for the duration of cycle 2. In particular it shows an anomalous transient decline in $m_{(J-K)}$, from JD $2449401$ to $2449440$, simultaneous with the 1994 state B$\to$A change, associated type II X-ray outburst and known NS accretion disc. Gaps in the data span 57(182) days before(after). Such a run of points at a low $m_{(J-K)}$ relative to the typical value of $0.52\pm0.05$ is unique in the 14 year dataset.

This large change in near IR fluxes cannot be due to the underlying B star, and the required increase in $T_{disc}$ while the overall disc flux is decreasing makes little sense in terms of normal models of disc emission. Another possibility is that being simultaneous with a state change, transient emission from residual low density material at radii between the new and old truncation radii may alter the global IR properties. However, such emission, assuming $T\leq T_{disc}$, is expected to have redder colours than the optically thick disc bulk \cite{dougherty1994} and so cannot be responsible.

An additional transient source of emission with a relatively blue colour index is thus suggested, the additional flux compensated for by the simultaneous reduction in Be disc emission associated with the state $B\to A$ change. If so an intriguing possibility is that this is emission from the known accretion disc around the NS. Limits can be found for the additional flux in $m_J$, the primary uncertainty being the Be disc flux. Based purely upon the colours, even if none of the additional flux is present in $m_K$ the 0.1 magnitude colour shortfall requires 10\% additional flux in $m_J$. Alternatively if we assume the Be star to have faded fully to a typical early state A luminosity then the total system magnitude excluding any accretion disc is $m_K^{Be}=7.45\pm0.10$ (Figure \ref{fig:a0535correlation1}) giving $m_J^{Be}=7.97\pm0.11$ (in excellent agreement with the value in Table \ref{tab:fluxquant}) assuming a typical $m^{Be}_{(J-K)}$ of $0.52\pm0.05$. The observed $m_J=7.80\pm0.07$ therefore yields a maximum excess of $0.17\pm0.13$ in $m_J$. Thus a 10-20\% excess in $m_J$ with a bluer $m_{(J-K)}$ than the dominant circumstellar emission is found.

Approximating the radius of the accretion disc to the NS Roche lobe radius at periastron and adopting the canonical $M_{NS}=1.4M_\odot$ and $M_*=26M_\odot$ \cite{vacca1996} yields $r_{L(NS)}=3.45\times10^{10}m$ \cite{eggleton}. Such a disc has 22\% of the emitting area of the Be disc in state A, which in turn emits 44\% of the J band flux. If we further assume optically thick emission and temperature $T_{AD}=T_{Be disc}$ an additional 9.7\% flux in $m_J$ is predicted. Therefore the estimate of a 10-20\% excess in $m_J$, combined with the bluer $m_{(J-K)}$, strongly suggests a significantly higher $T_{AD}$ and, possibly, a smaller emitting region.

The simultaneity of this unique set of photometric data within the 14 year dataset with an independent signature of an accretion disc in the system is noteworthy. The transient existence of an accretion disc in A0535+26 is thus supported.

\section{Disc expansion and Inclination}
\label{section:inclination}

\begin{figure}
\psfig{file=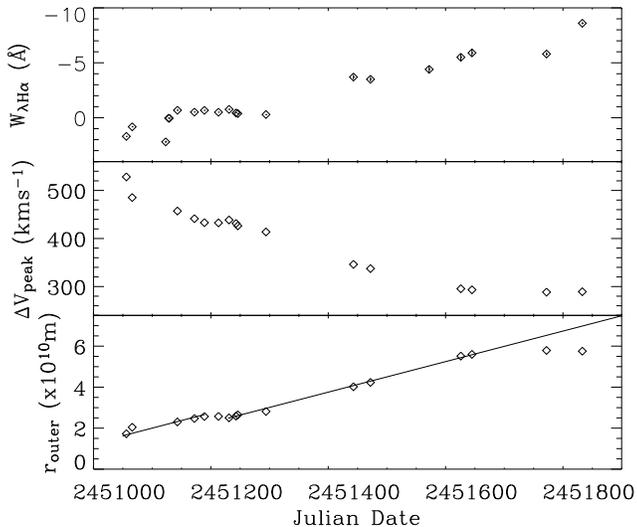,angle=0,width=3.3in}
\caption{\ha~line parameters following the 1998 disc-loss episode. Lower panel assumes $i=36.1^\circ$ and $M_*=26M_\odot$.}
\label{fig:discgrowth}
\end{figure}

Spectra obtained following the 1998 disc-loss episode provide a valuable glimpse of a reforming disc.  Figure~\ref{fig:discgrowth} plots \ewha, the separation of the peaks of the \ha~profile \deltav, and the derived disc radius $r_{outer}$. The fit to the data in the lower panel shows that $r_{outer}$ can be modelled as increasing linearly at $824ms^{-1}$ with a break around JD 2451200. During this break, disc expansion halts and \ewha~ remains constant; this presumably represents a hiatus in disc reformation. Such a highly subsonic expansion velocity is in good agreement with standard models of viscous Keplerian discs. 

Following disc-loss in 4U~0115+63, disc expansion continues until truncation becomes effective \cite{iggy2001}. In A0535+26 this can be seen to have occurred near JD 2451650; $m_K$ shows state A (Figure \ref{fig:a0535jhk}) and thus truncation at the 6:1 resonance (though note a brightening to state B and 5:1 shortly afterwards).

Assuming \ha~emission to the outer edge of a rotating disc, the Doppler displacement $v_{obs}$ of each peak of the resultant two peaked line profile represents the line of sight component of the Keplerian velocity at $r_{outer}$, whilst the resonant truncation paradigm informs us of the orbital period $p$ at this radius. Stee and Bittar \shortcite{stee2001} found that for typical isolated Be stars the \ha~emitting region was bounded at $18R_*$. As confirmed below, for all reasonable values of $M_*$ the disc is fully within this region in both states A and B.

The product of the orbital velocity and period yields $r_{outer}$

\begin{equation}
2\pi r_{outer} = \frac{v_{obs}}{sin i}p
\label{eq:getr}
\end{equation}

whilst for a particle in a circular orbit

\begin{equation}
r = \frac{sin^2i}{v_{obs}^2} G M_* 
\label{eq:r}
\end{equation}

\begin{figure}
\psfig{file=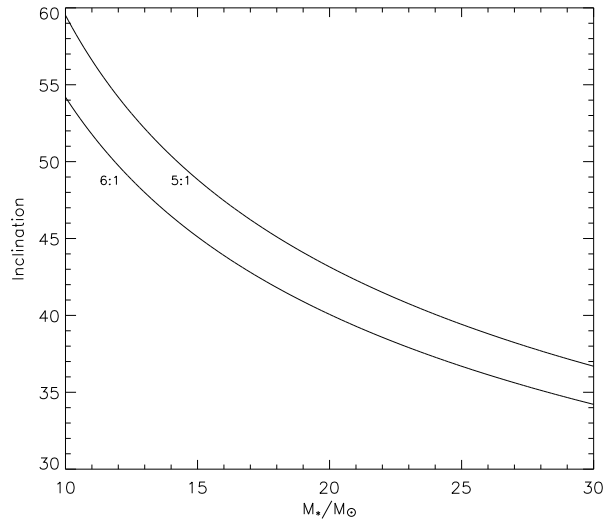,angle=0,width=3.3in}
\caption{$i$ vs. $M_*/M_\odot$ for two identifications of the truncation resonance in state A. Note 6:1 is favoured by theory and observation, discussed in Section \ref{section:IRquant}.}
\label{fig:mandi}
\end{figure}

Eliminating $r$ then gives a measure of $i$ as a function of $M_*$.

\begin{equation}
i = sin^{-1}\left(v_{obs}\left[\frac{p}{2\pi GM_*}\right]^{1/3}\right)
\label{eq:mandi}
\end{equation}

which is plotted in Figure \ref{fig:mandi} for all reasonable stellar masses. The case of the state A truncation at JD 2451650 described above is used, and \ha~\deltav=2$v_{obs}$~is measured at $288.9\pm2$\kms, in excellent agreement with the $275\pm20$\kms measured during the 1994 state A. For the reasons discussed in Section \ref{section:periods} the apparent NS orbital period relative to the disc is $P_{beat}=103$ days, though this almost no effect on the result. Section \ref{section:IRquant} established that state A is probably truncated at the 6:1 resonance giving $p=17.17$ days, though $i$ vs. $M_*$ is also plotted assuming a 5:1 commensurability yielding inclinations $\sim3^\circ$ larger.

Uncertainties stem largely from the uncertainty in $M_*$. The standard spectral classification O9.7IIIe requires a mass of $26\pm3M_\odot$ \cite{vacca1996}, yielding $36.1^\circ\pm1.7$; the less-favoured dwarf (V) classification ($M_*=22\pm3M_\odot$) yields $38.6^\circ\pm2.5$. These are at the low end of previous published results \cite{janot}.

Returning to Equation \ref{eq:getr}, $r_{outer}=(5.79\pm0.2)\times10^{10}m$ at a 6:1 truncation and $(6.68\pm0.25)\times10^{10}m$ at 5:1. Depending upon luminosity class and using $R_*$ from Vacca et al. \shortcite{vacca1996} 5:1 truncation corresponds to $r_{outer}=6.79R_*$($11.9R_*$) for luminosity class III(V) verifying the assumption of \ha~emission over the disc's full extent. This result clearly underscores the extent of truncation in BeXRBs relative to the discs of isolated Be stars.

Of course the rotating disc is not necessarily constrained to lie precisely in the equatorial plane of the star; Section \ref{section:periods} strongly suggests that it does not. However the constancy of the measured line velocities through the cycles requires that the tilt cannot be large ($\la10^\circ$) and so whilst this assumption should be born in mind the inclination should be accurate to approximately this tolerance.

\subsection{Determining $\omega$}

The disc-loss episode of 1998 provided a rare chance to obtain spectra free from circumstellar emission. Following Mazzali et al. \shortcite{mazzali} blue end \hei~lines were used to measure the stellar \vsini~due to their intrinsically narrow FWHMa, using photospheric spectra obtained shortly after disc-loss at the WHT. The preferred line, narrow with minimal emission, at 4713.20\AA~yields a \vsini~of $251\pm5$\kms, whilst the intrinsically broader $\lambda$4471,$\lambda$4922 lines give 254.2\kms~and 261.3\kms~respectively. Removing instrumental broadening, determined from arc lines to be 120\kms, yields a \vsini~of 220.4\kms~at $\lambda$4713, and 224 (232)\kms~for $\lambda$4471 ($\lambda$4922). 

Thus a value of \vsini=$225\pm10$\kms~is adopted. This compares with the published value of $250\pm15$\kms~ by Wang and Gies \shortcite{wanggies98}, who did not have the advantage of spectra uncontaminated by emission. A giant(dwarf) classification and thus $i=38.6^\circ\pm2.5$($i=36.1^\circ\pm1.7$) implies an equatorial rotational velocity of $361\pm25$($382\pm20$)\kms. 

The quantity $\omega$ gives the ratio of equatorial rotational velocity to 'critical' velocity, that of a particle in a circular orbit at the stellar equator. Porter \shortcite{porter1996} found that Be shell stars show $\omega=0.67\pm0.04$ and provided overwhelming evidence that this is true of Be stars in general. Depending upon a giant/dwarf classification, the masses and radii of Vacca et al. \shortcite{vacca1996} yield values for $v_{crit}$ of 582\kms(719\kms) for luminosity class III(V), giving $\omega=\frac{v}{v_{crit}}=0.62\pm0.05$($0.53\pm0.03$). This is consistent with a giant but excludes dwarves, supporting the O9.7IIIe classification.

\section{Periodicities and a global model}
\label{section:periods}

Whilst many period searches have been undertaken using photometry of A0535+26 (see Larionov et al.\shortcite{larionov2001}, henceforth L2001, for a review), only 103 and 1400 day periods repeatedly emerge at high significance whilst the orbital period is most notable by its absence. The 1400 day period is a manifestation of the cyclical photometric states described in this work, but the origin of the 103 day period is less clear.

\begin{figure}
\begin{center}
\centerline{\psfig{file=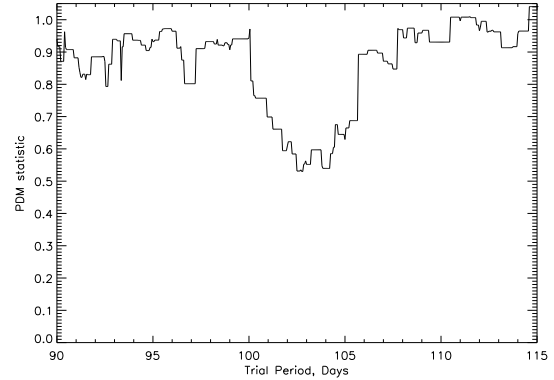,width=3in,angle=0}}
\caption{PDM periodogram of 54 $m_V$ data spanning 468 days from state A of cycle 2 (1994-5). Resolution is 0.05 days, using 6 PDM bins. Note strong detection at $103.1\pm0.1$ days and non-detection at the $\sim111$ day orbital period. Data from Lyuty and Za{\u \i}tseva (2000).}
\label{fig:pdm}
\end{center}
\end{figure}

\begin{figure}
\begin{center}
\centerline{\psfig{file=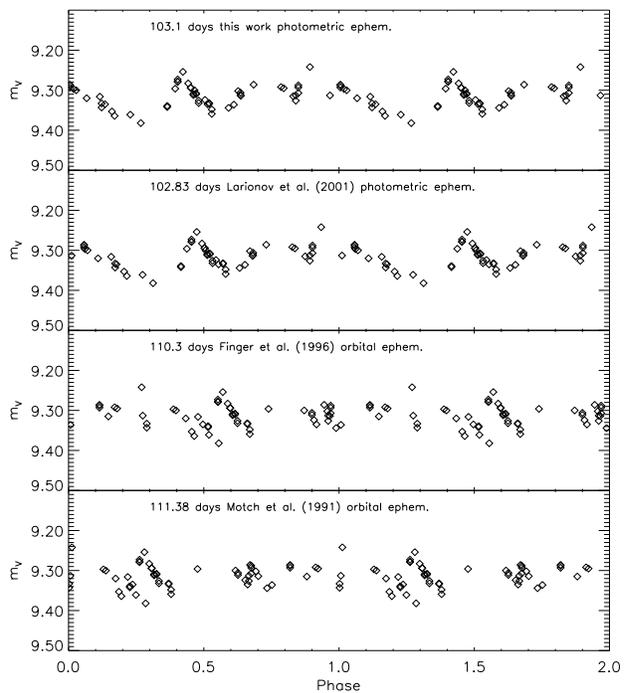,width=3.3in}}
\caption{$m_V$ photometry from state A of Cycle 2 (1994-5) folded at the detected beat period and two published orbital ephemerides. Data from Lyuty and Za{\u \i}tseva (2000).}
\label{fig:fold}
\end{center}
\end{figure}

Whilst the strongly shepherded structure of a truncated disc leads to highly stable IR photometric fluxes in all states, in the most strongly confined state (A) even $m_V$ demonstrates remarkably little random variation. This photometric stability is highly conducive to period searching. Thus 54 $m_V$ data spanning 468 days during state A of cycle 2 (1994-5) were searched for periods in the range 90-120 days using the PDM software \cite{pdm}. The resultant periodogram in Figure \ref{fig:pdm} again finds a strong $103.1\pm0.1$ day periodicity and nothing at $P_{orb}$. Folding using this period, the $102.83\pm0.15$ day period of L2001 and two published orbital periods, reveals the remarkable lightcurves in Figure \ref{fig:fold}.

While perturbations at the 103 day period result in disc flux changes by a factor of $\sim2$ in $m_V$, random fluctuations are $\leq 0.03$ magnitudes, and may in fact be due to observational error. Thus the luminosity, and thus presumably the disc structure, is determined {\em entirely} by the NS orbital phase. This may also be the cause of the larger range of $m_K$ in state A than state B (Figure \ref{fig:a0535correlation1}.

It is clearly of considerable relevance that the beat period of the orbital with the 1400 day period is 103 days. As L2001 suggest this is strongly suggestive of a 1400 day precessional period in one of the major flux producing system components. Adopting a range of $m_V=9.26\narrowto9.36$ from Figure \ref{fig:fold}, and using a photospheric $m_V=9.45\pm0.02$ from the 1998 disc-loss episode (Figure \ref{fig:a0535jhk} and Haigh et al. \shortcite{haigh99}) gives a range of $\sim2$ for the $P_{beat}$ fluctuations in disc emission. This variation is in severe conflict with the limit of a few percent to the {\em total} $m_V$ flux found by Lyuty and Za{\u i}tseva \shortcite{lyuty2000} for the possible contribution of an accretion disc. Furthermore the apparently transient nature of the NS accretion disc (Section \ref{section:disccolourchange}) argues against its involvement.

Therefore the precession must occur in the Be decretion disc, in a retrograde sense to the NS orbital motion in order to yield a NS-disc perturbation period shorter than $P_{orb}$. Indeed for a disc rotating in the same sense as the NS such retrograde precession is required \cite{wijers1998}. Simulations by Okazaki et al. \shortcite{okazaki2002} provide an example of how the disc's emission can then be modulated in $m_V$ at the 103 day period. However because their calculations of phase resolved disc flux assumed optically thin ff/bf emission, the probable optical thickness in $m_V$ during state A will alter the variations quantitatively but retain the periodicity. In addition to confirming lack of variability at $P_{orb}$, the apparent perturbations occurring at 103 days suggests that NS-disc interactions occur at this shorter period rather than $P_{orb}$. For this reason a 103 day period is used as the 'apparent' orbital period $P_{beat}$ in the flux modelling of Figure \ref{fig:cavity}.

Section \ref{section:cycles} showed how A0535+26's observation parameters, most notably the truncation radius, have in recent years varied in well defined cycles of mean duration $1590\pm180$ days. Using the detected $103.1\pm0.1$ day period as a beat period of $P_{orb}=111.0\pm0.5$ with a third period gives $1450\pm90$ days, consistent with the mean cycle duration; this equality is strongly suggestive of a common origin. The cycle period derived from measurements of $P_{beat}$ and $P_{orb}$ is probably more accurate than simply timing photometric state changes; the larger value found from cycle durations appears largely due to delayed increases in truncation radius as the Be disc declined around the 1996 minimum.

The $1590$ day cycles of truncation radius are explicable as arising from geometric variations in the strength of tidal torques as a function of precession angle. A possible clue to such continuous tidal variation is the trend from square-wave to quasi-sinusoidal variation at $\sim$1500 days towards shorter continuum wavelengths (compare $m_V$ to $m_K$ in Figure \ref{fig:a0535jhk} through cycles 1 and 2). Whilst $m_K$ represents emission from the surface of a well defined disc $m_V$, in general only partially optically thick, is sensitive to disc density and therefore reflects the continuously changing effect of tidal forces upon mass distribution in the disc.

A constant precession of the decretion disc - as opposed to transient precession prior to X-ray activity - is therefore the main suggested alteration to the model of 4U~0115+63 put forward by Negueruela et al. \shortcite{neg2001}.

The identification of specific truncation resonances allows constraints to be placed on the viscosity parameter $\alpha$ by comparison with the simulations and theoretical work of Okazaki and Negueruela \shortcite{natural2001}. The geometrically varying tidal forces will most closely match these theoretical predictions when the Be disc and NS orbit are most closely coplanar. At this time such forces will also be strongest and thus maximum truncation will occur; therefore state A, truncated probably at 6:1, occurs in the most coplanar phase. Thus for stability to occur at this truncation radius suggests the constraint $\alpha\leq0.013-0.038$ \cite{natural2001}.

\subsection{A general model for BeXRBs?}

The processes described in this work are fundamentally an extension of the theories put forward regarding the BeXRB 4U~0115+63 \cite{neg2001,iggy2001}. However, photometry of the BeXRB 4U~1145-619 presented by Stevens et al. \shortcite{stevens1997} clearly shows X-ray outbursts occurring during a fading in $m_V$ after a spell at a constant luminosity, an anticorrelation apparently identical to that described for A0535+26 in Section \ref{section:temporal}.

Approximately 4 yr periods of disc growth/dissipation strikingly similar to those reported here for A0535+262 have been reported for several other BeXRBs. Reig et al. \shortcite{reig2001} present photometry and spectroscopy of LS992/RX J0812.4-3114 showing a $\sim4.3$ year cycle; following disc-loss IR excess rapidly builds until the onset of X-ray activity, with IR excess fading slightly as Balmer emission peaks - this is identical to the activity cycles of A0535+26 seen in Figures \ref{fig:a0535jhk} and \ref{fig:a0535correlation1}.

The dataset on which the model for 4U~0115+63 (Section \ref{section:intro}, Negueruela et al. \shortcite{neg2001}) is based also shows near-identical variation of Balmer lines, IR excess and X-ray activity over 3-5 year cycles. 'Quasi-cyclic Be star envelope variations' have been invoked to explain the 4 year modulation in the amplitude of the periodic radio outbursts in LSI+61$^\circ$303 \cite{gregory1989}, while Hummel \shortcite{hummel1998} reported shell-Be-shell transitions in $\gamma~Cas$ at $4\pm0.4$ years which he attributed to a precessing circumstellar disc. Clark et al. \shortcite{clark2001} also show that X Per has exhibited disc-loss and disc 'low states' separated by intervals of 5 and 6 years.

If disc precession is the root cause of all these cycles, the similarity of cyclical period in systems with disparate orbital periods and inclinations is intriguing. While the disc is probably ousted from the equatorial plane by NS interactions, attribution of the {\em precession} to the asymetric potential of the oblate OB star is an appealing explanation, as the primaries of BeXRBs are known to occupy only a small range in $\omega$ \cite{porter1996} and $M_*$ \cite{neg1998} and thus should possess almost identical gravitational properties.

Behaviour similar to that observed in A0535+26, LS992 and 4U0115+63 is probably restricted to systems of moderate eccentricity causing the disc to be truncated at a low resonance (i.e. 4:1) so that changes between adjacent resonances cause observable flux changes. The degree of coplanarity of NS orbit and Be equator will clearly be an influential parameter profoundly affecting the nature of the tidal interactions. Orientation on the sky is also doubtless important for understanding different systems; it seems likely that the Be disc of 4U~0115+63 intercepts the line-of-sight at a specific cyclical phase, causing transient shell lines in that object and not in A0535+26. Clearly further multiwaveband studies of other BeXRB systems will demonstrate how general or indeed accurate this model is, particularly as differences emerge. 

\section{Summary}

The primary findings of this study are the detection of quantised disc fluxes strongly supporting the resonant truncation paradigm, association of X-ray outbursts with reductions in truncation resonance, and the r{\^ o}le of a $\sim$1500 day Be disc precession period in governing this system behaviour. Optical observations have revealed variability at the 'effective' $P_{orb}$, relative to the precessing Be disc, and represent Be disc perturbations by the NS. The standard luminosity classification, III, is supported.

This greater understanding of the A0535+26 system appears to be applicable to BeXRBs in general, perhaps allowing deeper insights into Be discs and quasi-Keplerian discs in general.

\section{Acknowledgements}

The Carlos S\'anchez Telescope is operated at the Teide Observatory (Tenerife) by the Instituto de Astrof\'\i sica de Canarias. The ING Archive and Service Programmes have provided invaluable data. We acknowledge MMT data courtesy of Laycock, Grindlay \& Zhao. Simon Clark is thanked for numerous helpful discussions.

\bsp
\label{lastpage}

\newpage


\begin{thebibliography}{99}

\bibitem[\protect\citename{{Alonso} et~al.~}{1998}]{alonso98}{Alonso} A., {Arribas} S., and {Martinez-Roger} C.: 1998, \newblock {\em Astron. Astrophys. Suppl. Ser.} {\bf 131}, 209

\bibitem[\protect\citename{{Baratta} et~al.~}{1978}]{baratta1978}{Baratta} G.~B., {Viotti} R. and {Altamore} A.:1978, \newblock {\em Astron. Astrophys.} {\bf 65}, L21

\bibitem[\protect\citename{{Clark~}}{1997}]{clarkythesis}
{Clark}, J.~S.: 1997,\newblock {\em Ph.D.~Thesis, University of Southampton}

\bibitem[\protect\citename{Clark et~al.~}{1998}]{clark98}Clark J., Tarasov A., Steele I., Coe M., Roche P., Shrader C., Buckley D., Larionov V., Larionova L., Lyuty V., Zaitseva G., Grunsfeld J., Fabregat J., and Parise R.: 1998,\newblock {\em Mon. Not. R. Astron. Soc.} {\bf 294}, 165

\bibitem[\protect\citename{{Clark} et~al.~}{1999}]{clarklyuty1999}{Clark} J.~S., {Lyuty} V.~M., {Zaitseva} G.~V., {Larionov} V.~M., {Larionova} L.~V., {Finger} M., {Tarasov} A.~E., {Roche} P., and {Coe} M.~J.: 1999,\newblock {\em Mon. Not. R. Astron. Soc.} {\bf 302}, 167

\bibitem[\protect\citename{{Clark} et~al.}{2001}]{clark2001}{Clark} J.~S., {Tarasov} A.~E., {Okazaki} A.~T., {Roche} P., and {Lyuty} V.~M.: 2001,
\newblock {\em Astron. Astrophys.} {\bf 380}, 615

\bibitem[\protect\citename{{de Loore} et~al.~}{1984}]{deloore1984}{de Loore} C., {Giovannelli} F., {van Dessel} E.~L., {Bartolini} C., {Burger} M., {Ferrari-Toniolo} M., {Giangrande} A., {Guarnieri} A., {Hellings} P., {Hensberge} H., {Persi} P., {Piccioni} A., and {van Diest} H.: 1984,
\newblock {\em Astron. Astrophys.} {\bf 141}, 279

\bibitem[\protect\citename{{Dougherty} et~al.~}{1994}]{dougherty1994}{Dougherty} S.~M., {Waters} L.~B.~F.~M., {Burki} G., {Cote} J., {Cramer}  N., {van Kerkwijk} M.~H., and {Taylor} A.~R.: 1994,
\newblock {\em Astron. Astrophys.} {\bf 290}, 609

\bibitem[\protect\citename{{Eggleton}~}{1983}]{eggleton}{Eggleton} P.: 1983,
\newblock {\em Astrophys. J.} {\bf 268}, 368

\bibitem[\protect\citename{{Finger} et~al.~}{1996}]{finger1996}{Finger} M.~H., {Wilson} R.~B., and {Harmon} B.~A.: 1996,
\newblock {\em Astrophys. J.} {\bf 459}, 288

\bibitem[\protect\citename{{Giovannelli} and {Ziolkowski}~}{1990}]{giovannelli1990}{Giovannelli} F., {Ziolkowski} J.:1990,
\newblock {\em Acta Astron.} {\bf 40}, 95

\bibitem[\protect\citename{{Giovannelli} et~al.~}{1999}]{iauc7293}{Giovannelli} F., {Sabau-Graziati} L., {Bernabei} S., and {Galleti} S.: 1999,
\newblock {\em IAU Circ.} 7293

\bibitem[\protect\citename{{Gregory} et~al.~}{1989}]{gregory1989}{Gregory} P.~C., {Xu} H., {Backhouse} C.~J., and {Reid} A.: 1989,
\newblock {\em Astrophys. J.} {\bf 339}, 1054

\bibitem[\protect\citename{Haigh et~al.~}{1999}]{haigh99}Haigh N.~J., Coe M.~J., Steele I.~A., and Fabregat J.: 1999,
\newblock {\em Mon. Not. R. Astron. Soc.} {\bf 310}, L21

\bibitem[\protect\citename{{Haigh~}}{~2002}]{mythesis}{Haigh} N.~J.: 2002,
\newblock {\em Ph.D.~Thesis, University of Southampton}

\bibitem[\protect\citename{{Hummel}}{1998}]{hummel1998}{Hummel} W.: 1998,
\newblock {\em Astron. Astrophys.} {\bf 330}, 243

\bibitem[\protect\citename{{Janot-Pacheco} et~al.}{1987}]{janot}{Janot-Pacheco} E., {Motch} C., and {Mouchet} M.: 1987,
\newblock {\em Astron. Astrophys.} {\bf 177}, 91

\bibitem[\protect\citename{{Larionov} et~al.~}{2001}]{larionov2001}{Larionov} V., {Lyuty} V.~M., and {Zaitseva} G.~V.: 2001,
\newblock {\em Astron. Astrophys.} {\bf 378}, 837

\bibitem[\protect\citename{{Li}~}{1997}]{li97}{Li} X-D.: 1997,
\newblock {\em Astrophys. J.} {\bf 476}, 278

\bibitem[\protect\citename{{Lyuty} et~al.~}{1989}]{lyuty1989}{Lyuty} V.~M., {Zaitseva} G.~V., and {Latysheva} I.~D.: 1989,
\newblock {\em Soviet Astronomy Letters} {\bf 15}, 182

\bibitem[\protect\citename{{Lyuty} and {Za{\u \i}tseva}~}{2000}]{lyuty2000}{Lyuty} V.~M. and {Za{\u i}tseva} G.~V.: 2000,
\newblock {\em Astronomy Letters} {\bf 26}, 9

\bibitem[\protect\citename{{Manfroid}~}{1993}]{manfroid}{Manfroid} J.: 1993,
\newblock {\em Astron. Astrophys.} {\bf 271}, 714

\bibitem[\protect\citename{{Mazzali} et~al.~}{1996}]{mazzali}{Mazzali} P.~A., {Lennon} D.~J., {Pasian} F., {Marconi} G., {Baade} D.,  and {Castellani} V.: 1996,
\newblock {\em Astron. Astrophys.} {\bf 316}, 173

\bibitem[\protect\citename{{Motch} et~al.~}{1991}]{motch1991}{Motch} C., {Stella} L., {Janot-Pacheco} E., and {Mouchet} M.: 1991,
\newblock {\em Astrophys. J.} {\bf 369}, 490

\bibitem[\protect\citename{{Negueruela}}~{1998}]{neg1998}{Negueruela} I.: 1998,
\newblock {\em Astron. Astrophys.} {\bf 338}, 505

\bibitem[\protect\citename{{Negueruela} and {Okazaki}~}{2001}]{iggy2001}{Negueruela} I. and {Okazaki} A.~T.: 2001,\newblock {\em Astron. Astrophys.} {\bf 369}, 108

\bibitem[\protect\citename{{Negueruela} et~al.~}{2001}]{neg2001}{Negueruela} I., {Okazaki} A.~T., {Fabregat} J., {Coe} M.~J., {Munari} U., and {Tomov} T.: 2001,\newblock {\em Astron. Astrophys.} {\bf 369}, 117

\bibitem[\protect\citename{{Okazaki} and {Negueruela}~}{2001}]{natural2001}{Okazaki} A.~T. and {Negueruela} I.: 2001,\newblock {\em Astron. Astrophys.} {\bf 377}, 161

\bibitem[\protect\citename{{Okazaki} et~al.~}{2002}]{okazaki2002}{Okazaki} A.~T., {Bate} M.~R., {Ogilvie} G.~I. and {Pringle} J.~E.: 2002,\newblock {\em astro-ph/0208288}

\bibitem[\protect\citename{{Persi} et~al.~}{1979}]{persi1979}{Persi} P., {Ferrari Toniolo} M., and {Spada} G.: 1979,
\newblock in {\em IAU Symp. 83: Mass Loss and Evolution of O-Type Stars}, 139

\bibitem[\protect\citename{{Piccioni} et~al.~}{1998}]{piccioni1998}{Piccioni} A., {Bartolini} C., {Bernabei} S. {et al.}: 1998,
\newblock in {\em Vulcano workshop 1998: Frontier objects in Astrophysics and Particle physics.}

\bibitem[\protect\citename{{Porter}~}{1996}]{porter1996}{Porter} J.~M.: 1996,
\newblock {\em Mon. Not. R. Astron. Soc.} {\bf 280}, L31

\bibitem[\protect\citename{{Reig} et~al.~}{2001}]{reig2001}{Reig} P.,{Negueruela} I.,{Buckley} D.~A.~H.,{Coe} M.~J.,{Fabregat} J.,{Haigh} N.~J.:2001,
\newblock {\em Astron. Astrophys.} {\bf 367}, 266

\bibitem[\protect\citename{{Rivinius} et~al.~}{2001}]{rivinius2001}{Rivinius} T., {Baade} D., {{\v S}tefl} S., 
and {Maintz} M.: 2001,\newblock {\em Astron. Astrophys.} {\bf 379}, 257

\bibitem[\protect\citename{{Stee} et~al.~}{1998}]{stee1998}{Stee} P., {Vakili} F., {Bonneau} D., and {Mourard} D.: 1998,\newblock {\em Astron. Astrophys.} {\bf 332}, 268

\bibitem[\protect\citename{{Stee} and {Bittar}}{2001}]{stee2001}{Stee} P. and {Bittar} J.: 2001,
\newblock {\em Astron. Astrophys.} {\bf 367}, 532

\bibitem[\protect\citename{Steele et~al.~}{1998}]{steele98}Steele I., Negueruela I., Coe M., and Roche P.: 1998,
\newblock {\em Astron. Astrophys.} {\bf 297}, L5

\bibitem[\protect\citename{{Stella} et~al.~}{1986}]{stella1986}{Stella} L., {White} N.~E., and {Rosner} R.: 1986,
\newblock {\em Astrophys. J.} {\bf 308}, 669

\bibitem[\protect\citename{{Stellingwerf}~}{1978}]{pdm}Stellingwerf R. F.: 1978, \newblock {\em Astrophys. J.} {\bf 224}, 953

\bibitem[\protect\citename{{Stevens} et~al.~}{1997}]{stevens1997}{Stevens} J.~B., {Reig} P., {Coe} M.~J., {Buckley} D.~A.~H., {Fabregat} J., and {Steele} I.~A.: 1997,\newblock {\em Mon. Not. R. Astron. Soc.} {\bf 288}, 988

\bibitem[\protect\citename{{Vacca} et~al.~}{1996}]{vacca1996}{Vacca} W.~D., {Garmany} C.~D., and {Shull} J.~M.: 1996,
\newblock {\em Astrophys. J.} {\bf 460}, 914

\bibitem[\protect\citename{{Vo{\u \i}khanskaya}}{1980}]{vo1980}{Vo{\u \i}khanskaya}, N.~F.: 1980,
\newblock {\em Sov. Astron. Lett.} {\bf 6}, 305

\bibitem[\protect\citename{{Wang} and {Gies}}{1998}]{wanggies98}{Wang} Z.~X. and {Gies} D.~R.: 1998,
\newblock {\em Publ. Astron. Soc. Pac.} {\bf 110}, 1310

\bibitem[\protect\citename{{Wijers} and {Pringle}}{1999}]{wijers1998}{Wijers} R.~A.~M.~J. and {Pringle} J.~E.: 1999,
\newblock {\em Mon. Not. R. Astron. Soc.} {\bf 308}, 207

\end{thebibliography}
\end{document}